\input harvmac
\input epsf
\input rotate
\input xyv2


\def\la{\Lambda}
\def\ep{\epsilon}


\Title{\vbox{\baselineskip12pt\hbox{CAS-KITPC/ITP-011}\hbox{USTC-ICTS-07-21}}
}{Holographic Cosmological Constant and Dark Energy}

\centerline{Chao-Jun Feng }

\bigskip
\bigskip

\centerline{\it Institute of Theoretical Physics, Academia Sinica}
\medskip
\centerline{\it P.O. Box 2735, Beijing 100080, China}
\bigskip
\centerline{\it and}
\bigskip
\centerline{\it Interdisciplinary Center of Theoretical Studies}
\medskip
\centerline{\it USTC, Hefei, Anhui 230026, China}
\bigskip
\centerline {fengcj@itp.ac.cn}

\bigskip

\bigskip

\noindent A general holographic relation between UV and IR cutoff of
an effective field theory is proposed. Taking the IR cutoff relevant
to the dark energy as the Hubble scale, we find that the
cosmological constant is highly suppressed by a numerical factor and
the fine tuning problem seems alleviative. We also use different IR
cutoffs to study the case in which the universe is composed of
matter and dark energy.


\Date{Sept. 2007}

\nref\RiessCB{
  A.~G.~Riess {\it et al.}  [Supernova Search Team Collaboration],
  Astron.\ J.\  {\bf 116}, 1009 (1998)
  [arXiv:astro-ph/9805201].
}
\nref\PerlmutterNP{
  S.~Perlmutter {\it et al.}  [Supernova Cosmology Project Collaboration],
  Astrophys.\ J.\  {\bf 517}, 565 (1999)
  [arXiv:astro-ph/9812133].
}
\nref\SpergelHY{
  D.~N.~Spergel {\it et al.}  [WMAP Collaboration],
  Astrophys.\ J.\ Suppl.\  {\bf 170}, 377 (2007)
  [arXiv:astro-ph/0603449].
}
\nref\AdelmanMcCarthyWU{ For a recent date release see,
  J.~K.~Adelman-McCarthy {\it et al.}  [SDSS Collaboration],
  arXiv:0707.3413 [astro-ph].
}
\nref\WeinbergCP{
  S.~Weinberg,
  Rev.\ Mod.\ Phys.\  {\bf 61}, 1 (1989).
}
\nref\CarrollFY{
  S.~M.~Carroll,
  Living Rev.\ Rel.\  {\bf 4}, 1 (2001)
  [arXiv:astro-ph/0004075].
}
\nref\PolchinskiGY{
  J.~Polchinski,
  arXiv:hep-th/0603249.
}
\nref\BoussoGP{
  R.~Bousso,
  arXiv:0708.4231 [hep-th].
}
\nref\tHooftGX{
  G.~'t Hooft,
  arXiv:gr-qc/9310026.
}
\nref\SusskindVU{
  L.~Susskind,
  J.\ Math.\ Phys.\  {\bf 36}, 6377 (1995)
  [arXiv:hep-th/9409089].
}
\nref\BoussoJU{ For a review of holographic principle see,
  R.~Bousso,
  Rev.\ Mod.\ Phys.\  {\bf 74}, 825 (2002)
  [arXiv:hep-th/0203101].
}

\nref\BekensteinUR{
  J.~D.~Bekenstein,
  Phys.\ Rev.\  D {\bf 7}, 2333 (1973);
  J.~D.~Bekenstein,
  Phys.\ Rev.\  D {\bf 23}, 287 (1981).
}
\nref\CohenZX{
  A.~G.~Cohen, D.~B.~Kaplan and A.~E.~Nelson,
  Phys.\ Rev.\ Lett.\  {\bf 82}, 4971 (1999)
  [arXiv:hep-th/9803132].
}
\nref\LiRB{
  M.~Li,
  Phys.\ Lett.\  B {\bf 603}, 1 (2004)
  [arXiv:hep-th/0403127].
  Q.~G.~Huang and M.~Li,
  JCAP {\bf 0503}, 001 (2005)
  [arXiv:hep-th/0410095].
  Q.~G.~Huang and M.~Li,
  JCAP {\bf 0408}, 013 (2004)
  [arXiv:astro-ph/0404229].
  B.~Chen, M.~Li and Y.~Wang,
  Nucl.\ Phys.\  B {\bf 774}, 256 (2007)
  [arXiv:astro-ph/0611623].
  J.~f.~Zhang, X.~Zhang and H.~y.~Liu,
  arXiv:0708.3121 [hep-th].
}

\nref\HoravaTB{
  P.~Horava and D.~Minic,
  Phys.\ Rev.\ Lett.\  {\bf 85}, 1610 (2000)
  [arXiv:hep-th/0001145].
}
\nref\MyungRJ{
  Y.~S.~Myung,
  Phys.\ Lett.\  B {\bf 578}, 7 (2004)
  [arXiv:hep-th/0306180].
}

\nref\HorvatVN{
  R.~Horvat,
  Phys.\ Rev.\  D {\bf 70}, 087301 (2004)
  [arXiv:astro-ph/0404204].
}
\nref\SheikhJabbariBJ{
  M.~M.~Sheikh-Jabbari,
  Phys.\ Lett.\  B {\bf 642}, 119 (2006)
  [arXiv:hep-th/0605110].
}
\nref\PodolskyVG{
  D.~Podolsky and K.~Enqvist,
  arXiv:0704.0144 [hep-th].
}
\nref\WittenZK{
  E.~Witten,
  arXiv:hep-ph/0002297.
}
%

\newsec{Introduction}
Why cosmological constant observed today is so much smaller than the
Planck scale? This is one of the most important problems in modern
physics. In history, Einstein first introduced the cosmological
constant in his famous field equation to achieve a static universe
in 1917. After the discovery of the Hubble's law, the cosmological
constant was no longer needed because the universe is expanding.
Nowadays, the accelerating cosmic expansion first inferred from the
observations of distant type Ia supernovae \RiessCB\PerlmutterNP\
has strongly confirmed by some other independent observations, such
as the cosmic microwave background radiation (CMBR) \SpergelHY\ and
Sloan Digital Sky Survey (SDSS) \AdelmanMcCarthyWU, and the
cosmological constant returns back as a simplest candidate to
explain the acceleration of the universe in 1990's.

In particle physics, the cosmological constant naturally arises as
an energy density of the vacuum , which is evaluated by the sum of
zero-point energies of quantum fields with mass $m$  as follows
\eqn\qft{\rho_\la={1\over2}\int_0^\la{4\pi k^2 dk\over
(2\pi)^3}\sqrt{k^2+m^2}\approx{\la^4\over16\pi^2} ,}
where $\la\gg m$ is the UV cutoff. Usually the quantum field theory
is considered to be valid just below the Planck scale: $M_p\sim
10^{18}$GeV, where we used deduced Planck mass $M_p^{-2}=8\pi G$ for
convenience. If we pick up $\la=M_p$, we find that the energy
density of the vacuum in this case is estimated as $10^{70}$GeV$^4$,
which is about $10^{117}$ orders of magnitude larger than the
observation value $10^{-47}$GeV$^4$.  One may try to cancel it by
introducing counter terms, however, this requires a fine tuning to
adjust the energy density of the vacuum to the present energy
density of the universe (for a classic review see \WeinbergCP, for a
recent nice review see \CarrollFY, and for a recent discussion see
\PolchinskiGY\BoussoGP). It seems that the number of the independent
degrees of freedom of the quantum fields should not be very large
\tHooftGX\SusskindVU.

Holographic principle \BoussoJU\ regards black holes as the
maximally entropic objects of a given region and postulates that the
maximum entropy inside this region behaves non-extensively, growing
only as its surface area. Hence the number of independent degrees of
freedom is bounded by the surface area in Planck units, so an
effective field theory with UV cutoff $\la$ in a box with size $L$
will make sense if it satisfies the Bekenstein entropy bound
\BekensteinUR\
\eqn\beb{(L\la)^3\leq S_{BH}=\pi L^2M_{pl}^2 \, ,}
where $M_{pl}^{-2}\equiv G $ is the Planck mass and $S_{BH}$ is the
entropy of a black hole of radius $L$ which acts as an IR cutoff.
Cohen and collaborators \CohenZX\ suggested that the total energy in
a region of size $L$ should not exceed the mass of a black hole of
the same size
\eqn\cohen{L^3\la^4\leq LM_p^2 \, ,}
which can be simply rewritten as
\eqn\cohens{(L\la)^4\leq L^2M_p^2 \, .}
This bound is much more stringent than the bound \beb: when equation
\cohens\ is near saturation, the entropy of the quantum field is
\eqn\entropy{S_{max}\approx S^{3/4}_{BH} \, .}

Since we have limited knowledge about the holographic principle and
we have not even know whether the holographic principle is right or
not because we have only a few examples to realize it. The only
successful example to my knowledge is the AdS/CFT correspondence.
Mostly, one believes the holographic principle is right because it
does not conflict with any observations so far. As a result we can
not claim whether the bounds mentioned above as a consequence of the
holographic principle is correct or not, and it may be too stringent
or too loose due to some unknown reasons or some underlying theory.
In section $2$ we postulate a general bound which  provides a
mechanism to derive a very small vacuum energy from the principle of
holography.

When the matter presents in the universe, the evolution of the dark
energy (in this note we shall use terms the cosmological constant
and the dark energy exchangeably) is sensitive to the chosen of the
IR cutoff. When we take the event horizon as the IR cutoff, the
result is very similar to the case studied in ref. \LiRB\ up to some
corrections. As long as the vacuum dominates the energy density in
the later time, it should be small as we discussed in section 2.

This paper is organized as follows. In section 2 we postulate a
general relation between the UV an IR cutoff and the smallness of
the cosmological constant shall be explained. In the next section we
use three different IR cutoffs to study the evolution and the
equation of state of the dark energy. In the final section we will
give some discussions.

\newsec{General bound and the cosmological constant}

In this note we postulate a general relation of UV and IR cutoff as
follows
\eqn\gen{(L\la)^n\leq L^2M_p^2 \, ,}
where $n$ is a dimensionless parameter that comes from some
underlying theory. When $n=3,4$ equation \gen\ is reduced to \beb\
and \cohens\ respectively, and we shall see that the final
consistent $n$ is slightly deviation from $4$ but without any fine
tuning. Of course we can not say anything about this unknown theory
yet, since we do not even know whether there really exists such a
theory or not, but we shall see that if the relation \gen\ is
correct it will provide a mechanism to derive a small cosmological
constant without any fine tuning. There are some works trying to
solve the cosmological constant problem from the holographic
principle, for instance see
\HoravaTB\MyungRJ\LiRB\HorvatVN\SheikhJabbariBJ\PodolskyVG, but they
do not consider the general case of the relation \gen\ between UV
and IR cutoff.

The largest $\la$ allowed here is the one saturating the inequality
\gen:
\eqn\sat{\la=L^{{2\over n}-1}M_p^{{2\over n}} \, .}
Then the energy density of the vacuum $\rho_{\la}\sim\la^4$ is
\eqn\vac{\rho_{\la}=3c^2L^{{8\over n}-4}M_p^{{8\over n}} \, ,}
where a numerical constant $3c^2$ is introduced in the above
equation for convenience. From \qft\ one can see the value of $c^2$
is naturally neither very large nor very small. Of course there is
also a constant in the equation \sat, but naturally such a constant
could not be very large or very small and it will not affect the
final conclusion since one can absorb this constant into $c^2$.

The dynamics of the universe is described by the Einstein field
equations. The observations indicate that the universe is
homogeneity and isotropy on large scales and the generic metric
respecting these symmetries is the Friedmann- Robertson-Walker (FRW)
metric given by
\eqn\frwm{ds^2=-dt^2+a^2(t)\left[{dr^2\over1-Kr^2}+r^2(d\theta^2+sin^2\theta
d\phi^2)\right] \, ,}
where $a(t)$ is scale factor with cosmic time $t$. The coordinates
$r$, $\theta$ and $\phi$ are known as comoving coordinates. A freely
moving particle comes to rest in these coordinates. The constant $K$
in the metric \frwm\ describes the geometry of the spatial section
of space time, and $K=+1, 0, -1$ corresponds to closed, flat and
open universe respectively.

Consider an ideal perfect fluid with energy density $\rho$ and
pressure $p$ as the source of the energy momentum tensor and solve
the Einstein equation with the metric \frwm, we find the famous
Friedmann equations
\eqn\friequa{H^2={\rho\over3M_p^2}-{K\over a^2}}
and
\eqn\frieII{\dot H=-{\rho+p\over2M_p^2}+{K\over a^2}}
where $H\equiv\dot a/a$ is the Hubble constant. Since observations
indicate the universe is flat, i.e. the critical energy density is
almost equal to $1$, we will only consider the flat case $K=0$ in
the following. In fact \frieII\ can be derived with the help of the
continuity equations respecting the conservation of the energy
momentum as the consequence of the Bianchi identities:
\eqn\cebian{\dot\rho+3H(\rho+p)=0 .}

To indicate the application of the relation \gen, we consider such a
situation in which we assumed that the IR cutoff is the Hubble scale
$H^{-1}$ and the vacuum dominates the universe:
\eqn\fried{3M_p^2H^2=3c^2H^{4-{8\over n}}M_p^{{8\over n}} \, ,}
which can be easily solved
\eqn\sol{\left({H\over M_p}\right)^{2-{8\over n}}={1\over c^2} \, .}
For a given $n$ ($n\neq4$) $H$ is a constant, so the universe is de
Sitter space. While $n=4$, then $c^2=1$ and $H$ could be any value
since $n=4$ is an unstable point, thus we can get very small value
of cosmological constant when $n$ is slightly deviation from $4$,
and there is not any fine tuning problem in such a difference, we
shall see this in the following. We would like to emphasize that $n$
and $c$ here are determined by some underlying reasons as a normal
number, by normal number we mean the number is not very large like
$10^{10}$ or small like $10^{-10}$, so there is no need for us to
adjust them to produced a small vacuum energy. If our postulation is
right, it should make it. Usually $c^2$ is not equal to $1$ , so the
energy density of the vacuum is
\eqn\vacum{\rho_{\la}=3c^2H^{4-{8\over n}}M_p^{{8\over
n}}=3(c^2)^{{n\over 4-n}}M_p^4}
where we have used \sol. If $c^2<1$ and $n<4$, the energy density
can be highly suppressed and much smaller than the Planck scale
energy density $M_p^4$ in the limit of $n\rightarrow4$. If $c^2>1$
and $n>4$, one can also get a very small energy density. In the
other case the value of energy density is ruled out by the
observations because it is too large.  To illustrate that $n$ has
the slight but not fine-tuning difference from $4$, we give a
concrete example in the following.

Assuming $c^2=0.1$, and if
\eqn\nva{n=n_{0.1}={4\over1-{\ln c^2\over 117\ln10}}\approx3.966 \,
,}
the energy density is roughly $10^{-117}M_p^4$. Let $n=4+\ep$, one
can see $\ep=-0.03$ in this example. In fact, such a difference
$|\ep|$ would not be a extreme small number, namely, a number like
$10^{-10}$, as long as $c^2$ is not very closed to $1$, so there is
no fine tuning problem here, one can see this property in Figure.~1.

\bigskip
{\vbox{{
        \nobreak
    \centerline{\epsfxsize=7cm \epsfbox{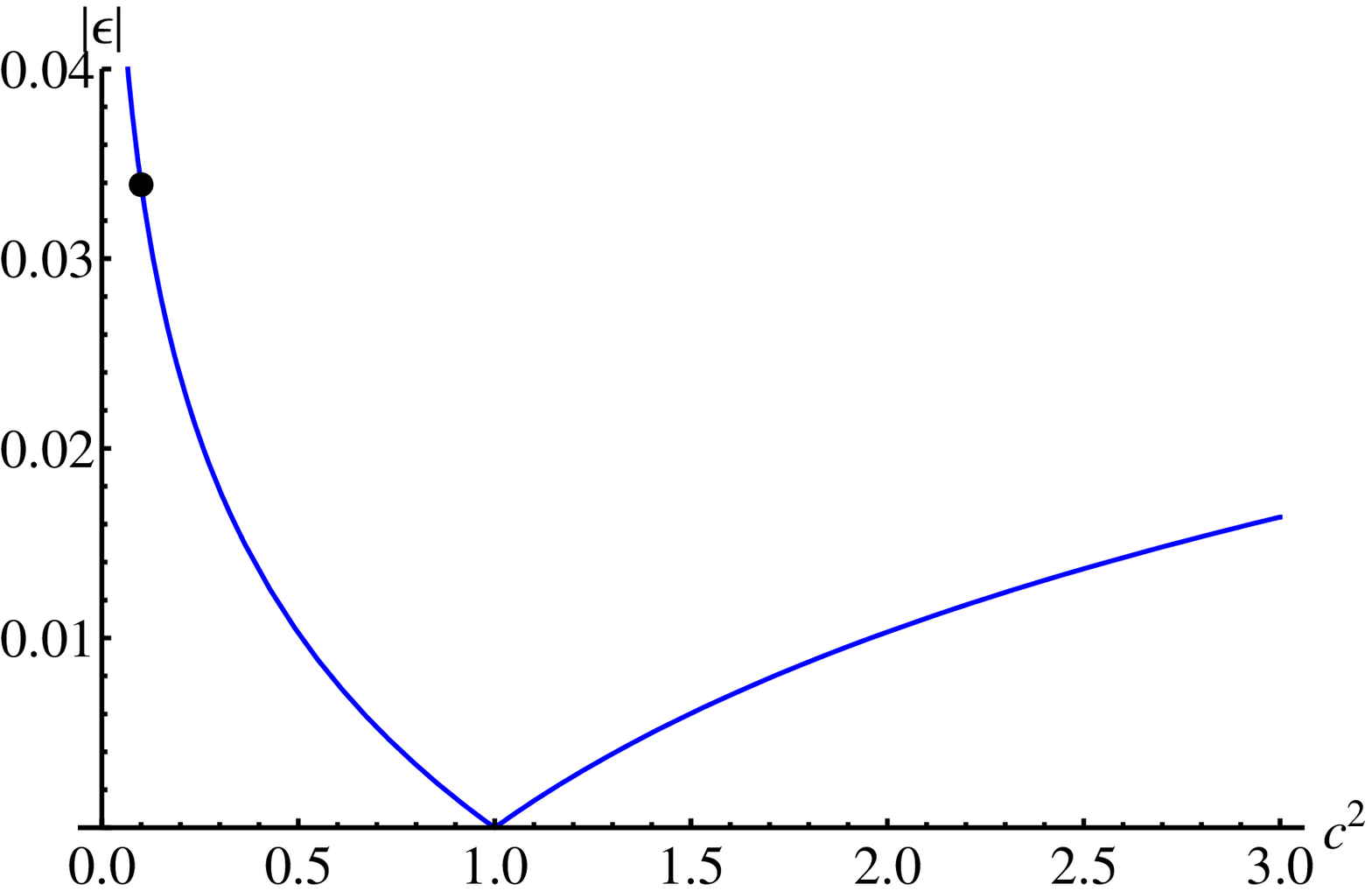}}
        \nobreak\bigskip
    {\raggedright\it \vbox{
{\bf Figure 1.} {\it The difference $|\ep|\equiv|n-4|$ vs. $c^2$.
The value of $c^2$ and $\ep$ on the curve will deduce a small energy
density, namely, $10^{-117}M_p^4$. The black point on the curve is
the point $c^2=0.1$, $|\ep|\approx0.03$.}
 }}}}
    \bigskip}

At first glance it seems that there is a fine tuning problem here:
for a given $c^2$ , one should adjust the value of $n$ to be very
closed to $n_{0.1}$ for example. But it is not the case, because we
do not need to adjust $c^2$ and $n$ in fact, the resulting energy
density is a consequence or a prediction of the theory rather than a
input. The figure above only indicates that the underlying theory
should contain a constant number $n$ whose value is near $4$ without
any fine tuning.

Let's have a look at equation \gen\
\eqn\gens{(L\la)^{4+\ep}\leq L^2M_p^2 \, .}
It seems that we are living in the fractal dimension spacetime
rather than $4$, if one regards the power of $L\la$ as the dimension
of the world we living due to the simple fact that the max entropy
in $4$ dimension spacetime is roughly $L^3\la^3$ and in $3$
dimension spacetime is $L^2\la^2$. It is amazing if this explanation
is correct because it is so counterintuitive. But we are forced to
reconsider the stability of the orbits of planets like the earth
because there are no stable solutions to keep the earth rounding the
sun, in other words the Inverse Square Law is not hold in high
dimensions. Things is also bad in low dimensional worlds because no
one can live in $2$ dimension space. Problems may be disappeared
here because of the fractal dimension.  Another possibility is that
in the following.

Take the correction term to the R. H. S. of \gens\ as
\eqn\genr{L^3\la^4\leq (L\la)^{-\ep}LM_p^2 .}
The factor $(L\la)^{-\ep}$ may come from the some unknown theory.
Since this factor has something to do with the cosmological constant
which can be considered as a consequence of quantum gravity
\WittenZK, one could guess this factor may come from the correction
of quantum gravity theory. Maybe these two possibilities are the
same thing, but there is no evidence here now.

\newsec{Dark energy with the presence of matter}

With matter present, the Friedmann equation reads
\eqn\friemat{3M_p^2H^2=\rho_m+\rho_\la \, ,}
where $\rho_m$, the energy density of matter, satisfies the
continuity equation:
\eqn\mattercon{\ln'\rho_m+3=0 \, .}
where prime denotes the derivative with respect to $\ln a$. The
dimensionless energy density of matter is defined as
$\Omega_m=\rho_m/(3M_p^2H^2)$, so it satisfies the following
equation
\eqn\dmated{\ln'\Omega_m=-3-2\ln'H \, ,}
where we have used the continuity equation \mattercon.

Define a dimensionless energy density
\eqn\megamat{\Omega_\la\equiv{\rho_\la\over
3M_p^2H^2}=c^2(LM_p)^{{8\over n}-4}H^{-2}M_p^2 .}
Then the Friedmann equation \friemat\ is
\eqn\fridl{\Omega_m+\Omega_\la=1 .}

Derivative the logarithm of \megamat\ with respect to $\ln a$ as
follows
\eqn\eqtres{\ln'\Omega_\la=-2\left[(2-{4\over
n})\ln'L+\ln'H\right]\approx-2\left[\ln'(LH)+{\ep\over4}\ln'L
\right]}
From \dmated, \fridl\ and \eqtres\ we can derive a equation of $H$
and $L$
\eqn\fineq{\ln'H+(1+{\ep\over4})\Omega_\la\ln'L+{3\over2}(1-\Omega_\la)=0}
and once we know the another relation of $L$ and $H$, we can solve
this equation.

Measuring $w$ as in $\rho_\la\sim a^{-3(1+w)} $, we have the index
$w$ given by
\eqn\eodw{w=-1-{1\over3}\left({d\ln\rho_\la\over d\ln
a}+{1\over2}{d^2\ln\rho_\la\over d(\ln a)^2}\ln a\right) \, ,}
up to the second order and the derivatives are taken at the present
time $a_0 = 1$. From \vac\ we find
\eqn\eqtoo{{d\ln\rho_\la\over d\ln a}={\dot\rho_\la\over
H\rho_\la}=\left({8\over n}-4\right){\dot L\over L
H}\approx-2(1+{\ep\over4})\ln'L}
thus we get the ration of pressure to energy density $w$ as
\eqn\eodwg{w\approx-1+{2\over3}(1+{\ep\over4})\ln'L .}
For the acceleration of the universe $w<-1/3$, the R. H. S. of
\eodwg\ should satisfy
\eqn\posis{(1+{\ep\over4})\ln'L<1 \, .}
and for a increasing of $\Omega_\la$ the R.H.S. of \eqtres\ should
be positive
\eqn\posi{\ln'(LH)+{\ep\over4}\ln'L<0 .}
From \posis\ one can see that, when the IR cutoff $L$ is smaller
than $1$, the universe is acceleration definitely and this vacuum
energy will eventually dominate the universe. This happens when we
regards the event horizon as a natural cutoff as the IR cutoff, and
Miao's work in \LiRB\ has already indicated such a character.

In the following we will study the property of the vacuum energy
with three different IR cutoffs : Hubble scale $H^{-1}$, particle
horizon $R_p$ and event horizon $R_h$, since these IR cutoffs
naturally arise when one studies the universe. The definition of
$R_p$ and $R_h$ is given by
\eqn\eh{R_p(t)=a(t)\int_0^t {dt'\over a(t')}}
\eqn\ehI{R_h(t)=a(t)\int_t^\infty {dt'\over a(t')}}

\subsec{Case1: $L=H^{-1}$}

In this case, the vacuum energy behaves almost like the matter,
which means it's equation of state is very similar to that of the
matter up to some corrections, and we find this correction will lead
to an evolution of $w$, but it will take a long time for $w$ to be
$-1$. In other words, the vacuum energy will not accelerate the
universe until it almost completely dominates the universe. Since
the calculation discussed above is straightforward, we simply give
the final result as follows.

From \eqtres\ and \fineq\  the equation of the energy density is
found to be
\eqn\omgeom{{\Omega_\la'\over\Omega_\la}=-{3\ep\over4}\left[{1-\Omega_\la\over1-(1+{\ep\over4})\Omega_\la}\right]
\, ,}
where $L=H^{-1}$ was used. When $\ep<0$ the dimensionless energy
density of the vacuum is increasing with time. This equation can be
solved easily as
\eqn\somgeom{\ln\Omega_\la+{\ep\over4}\ln(1-\Omega_\la)=-{3\ep\over4}\ln
a+x_0}
If we set $a_0 = 1$ at the present time, $x_0$ is equal to the
L.H.S. of \somgeom with $\Omega_\la$ replaced by $\Omega_\la^0$ ,
namely, $x_0=\ln\Omega_\la^0+{\ep\over4}\ln(1-\Omega_\la^0)$. As
time draws by, $\Omega_\la$ increases to $1$, the second term on the
L.H.S. of \somgeom\ is the important term, we find, for large $a$
\eqn\largea{\Omega_\la=1-e^{4x_0/\ep}a^{-3} .}
Since the universe is dominated by the dark energy for large $a$, we
have
\eqn\larar{\rho_\la\sim\rho_c={\rho_m\over1-\Omega_\la}={\rho_m^0a^{-3}\over1-\Omega_\la}}
Thus, using \largea\ in the above relation
\eqn\lararl{\rho_\la=e^{-4x_0/\ep}\rho_m^0}
which is too large compared with the observation value of
$\Omega_\la^0$, if we require a acceleration universe, namely
$\ep<0$.

For small $a$, matter dominates, the important term on the L.H.S. of
\somgeom\ is the first term, we find
\eqn\smalla{\Omega_\la=a^{-3\ep/4}e^{x_0}}
thus
\eqn\smallar{\rho_\la=\Omega_\la\rho_c=\Omega_\la\rho_m=e^{x_0}\rho_m^0a^{-3(1+\ep/4)}
,}
here $\ep$ is much smaller than $1$, so the evolution of the vacuum
energy is roughly $a^{-3}$ the same as the matter when $a$ is small.
In other words $w$ is almost zero when matter presents.

Up to the second order, the equation of sate is described by
\eqn\seodw{w={\ep\over4-(4+\ep)\Omega_\la^0}+{3\ep^2(1+\ep/4)\Omega_\la^0(1-\Omega_\la^0)\over32(1-(1+\ep/4)\Omega_\la^0)^3}z}
where we used $\ln a = -\ln(1 + z)\sim-z$.

Specifying to the case $\ep=-0.03$ and plugging the optional value
$\Omega_\la^0= 0.73$ into \seodw,
\eqn\sseodw{w=-0.027+7.9\times10^{-4}z.}
It seems that the Hubble scale is not a suitable IR cutoff.

\subsec{Case2: $L=R_p$}

If we take the particle horizon as the IR cutoff, the situation is
not much changed from the Hubble scale case. Since
\eqn\phe{\ln'L=\ln'R_p=1+{1\over R_pH }}
the equation of state from \eodwg\ is
\eqn\pheeos{w=-{1\over3}+{\ep\over6}+{2(1+{\ep\over4})\over3R_pH}}
which is larger than ($-1/3+\ep/6$). It seems that the universe is
hardly to accelerate in this case.

From \eqtres\ and \fineq\ we find
\eqn\eomphd{\ln'\Omega_\la=-2(1-\Omega_\la)\left[(1+{\ep\over4})(1+{1\over
R_pH})-{3\over2}\right] ,}
so the vacuum energy will dominates at later time if
\eqn\eomphdw{R_pH>2\left(1+{3\ep\over4}\right)}
It seems that the particle horizon is not a suitable IR cutoff
either.

\subsec{Case3: $L=R_h$}

In this case the situation is changed, namely we can get an
accelerating universe as follows and firstly one can simply see that
\eqn\ehem{\ln'L=\ln'R_h=1-{1\over R_hH }, }
where the minus sign in \ehem\ is the main difference from \phe\ ,
so the equation of state \pheeos\ is changed to be
\eqn\pheeos{w=-{1\over3}+{\ep\over6}-{2(1+{\ep\over4})\over3R_hH}}
which is smaller than ($-1/3+\ep/6$). It seems that the universe is
able to accelerate in this case. Here the term $\ep/6$ will slightly
change the value of $w$, and this is a correction to that in \LiRB.

From \eqtres\ and \fineq\ we find
\eqn\eomphdeh{\ln'\Omega_\la=-2(1-\Omega_\la)\left[(1+{\ep\over4})(1-{1\over
R_hH})-{3\over2}\right] ,}
so the vacuum energy will dominates if
\eqn\eomphdwe{{1\over R_pH}>-{1\over2}\left(1-{3\ep\over4}\right)}
which is always hold. Use the definition of $\Omega_\la$ in
\megamat, we find \eomphdeh\ becomes
\eqn\eomphdehb{\ln'\Omega_\la=-2(1-\Omega_\la)\left[(1+{\ep\over4})(1-{\sqrt{\Omega_\la}\over
c}(R_hM_p)^{\ep/4})-{3\over2}\right] ,}
which can not be solved analytically. The approximate solution when
$\ep$ is small will reduce to the result in \LiRB.  The
corresponding equation of state from \pheeos\ will be
\eqn\ceosis{w=-{1\over3}+{\ep\over6}-{2(1+{\ep\over4})\over3c}\sqrt{\Omega_\la}\,
\left(R_hM_p\right)^{\ep/4}.}
Where we have used \megamat. If $\ep=0$, the above equation is the
same as that in \LiRB, so terms containing $\ep$ are corrections
with slight effects.

When the vacuum completely dominates the universe at last, the
universe is a de Sitter space and the event horizon is roughly the
inverse of the Hubble constant at that time $t_0$, namely $R_h\sim
H^{-1}_0$. If we take today's value of the Hubble constant
$H_0^4\sim10^{-117}M_p^4$, the factor $\left(R_hM_p\right)^{\ep/4}$
in \ceosis\ is roughly $\sim0.60$ where we have used $\ep\sim-0.03$.
Taking the present value of $\Omega_\la^0=0.73$ and $c=0.5$ then
$w_0\approx-1.02$. Figure.~2 shows the relation between $w_0$ and
$c$, but notice that here $c$ as a constant is from some unknown
theory rather than adjusted.

\bigskip
{\vbox{{
        \nobreak
    \centerline{\epsfxsize=7cm \epsfbox{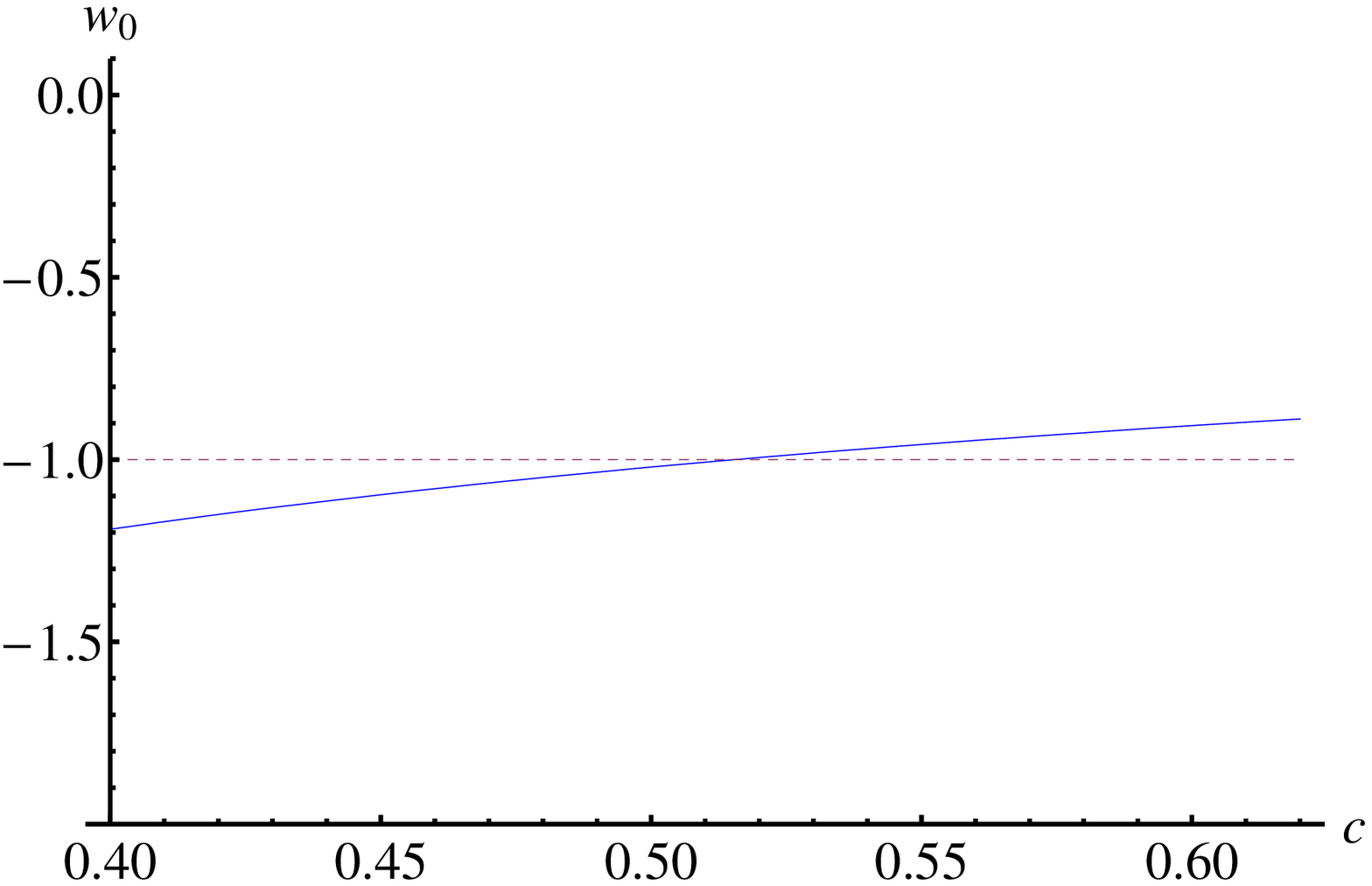}}
        \nobreak\bigskip
    {\raggedright\it \vbox{
{\bf Figure 2.} {\it The dash line denotes the line $w_0=-1$. The
solid line indicates $w_0$ vs. $c$ when $\ep=-0.03$ and the factor
$\left(R_hM_p\right)^{\ep/4}\sim0.6$.  }
 }}}}
    \bigskip}

\newsec{Discussions}

The application of  holographic principle discussed in the present
paper alleviates the cosmological constant problem. When the vacuum
dominates the universe, the energy density could be very small due
to the number of $n$ and $c^2$ come from some underlying reasons.
This provide a mechanism to explain why the cosmological constant is
so small. In other words, one can get a very small energy density
consistent with the observation value by this mechanism. We give an
example and argue that there is no fine tuning problem in this
mechanism. It should be emphasized that $n$ and $c^2$ are given
numbers rather than adjusted.

When matter presents as a component in the universe, this vacuum
energy play a role as the dark energy. The evolution of the dark
matter here is sensitive to the IR cutoff. We have used three
different cutoff and find the result is consistent with \LiRB,
namely the event horizon is a suitable IR cutoff for the energy
density to accelerate the universe.

At  first glance it seems that if $\ln'L$ is non-positive the
universe is definitely accelerating from \posis. However, if
$\ln'L<0$, it means there is a shrinking IR cutoff, so the cutoff
will be smaller and smaller as time draws by and we can see less and
less stars and galaxies. This is absurd. If $\ln'L=0$, it means
there is a universal IR cutoff of the universe, by universal we mean
the cutoff is independent of cosmic time, then the energy density of
the dark matter is a constant, namely $w=-1$, but there is no
evidence that we have such a universal IR cutoff. In a word, $\ln'L$
should be positive.

\bigskip
\noindent{\bf Acknowledgements}

We are grateful to Xian Gao, Wei Song, Yushu Song, Tower Wang, Yi
Wang, Wei Xue, and Xin Zhang for useful discussions.  The author
acknowledges Miao Li for a careful reading of the manuscript and
valuable suggestions.

\listrefs

\bye